\documentclass[preprint]{aastex}

\usepackage{natbib}
\usepackage{color}
\slugcomment{IPMU13-0012}

\shorttitle{Dark halos in WDM models and models with Long-Lived CHAMP}
\shortauthors{Kamada et al.}

\begin{document}


\title{Structure of Dark Matter Halos in Warm Dark Matter models and in models 
with Long-Lived Charged Massive Particles}


\author{Ayuki Kamada, Naoki Yoshida}
\affil{Kavli Institute for the Physics and Mathematics of the Universe, University of Tokyo, Kashiwa, Chiba 277-8583, Japan\\
Department of Physics, University of Tokyo, Tokyo 113-0033, Japan}

\email{ayuki.kamada@ipmu.jp}

\author{Kazunori Kohri}
\affil{Cosmology group, Theory Center, IPNS, KEK, and The Graduate University for Advanced Study (Sokendai), Tsukuba, 305-0801, Japan}

\and

\author{Tomo Takahashi}
\affil{Department of Physics, Saga University, Saga 840-8502, Japan}





\begin{abstract}

We study the formation of non-linear structures in Warm Dark Matter (WDM) models
and in a Long-Lived Charged Massive Particle (CHAMP) model.
CHAMPs with a decay lifetime of about $1 \,{\rm yr}$ induce characteristic 
suppression in the matter power spectrum at subgalactic scales 
through acoustic oscillations in the thermal background. 
We explore structure formation in such a model.
We also study three WDM models, where the dark matter particles are produced
through the following mechanisms:
i) WDM particles are produced in the thermal background and then kinematically 
decoupled; 
ii) WDM particles are fermions produced by the decay of thermal heavy bosons; 
and iii) WDM particles are produced by the decay of non-relativistic heavy particles.
We show that the linear matter power spectra for the three models are 
all characterised by the comoving Jeans scale at the matter-radiation equality.
Furthermore, we can also describe the linear matter power spectrum for the Long-Lived CHAMP model 
in terms of a suitably defined characteristic cut-off scale $k_{\rm Ch}$,
similarly to the WDM models.
We perform large cosmological $N$-body simulations to study the non-linear
growth of structures in these four models.
We compare the halo mass functions, the subhalo mass functions, and the radial 
distributions of subhalos in simulated Milky Way-size halos.
For the characteristic cut-off scale $k_{\rm cut} = 51\,h\,{\rm Mpc}^{-1}$, 
the subhalo abundance ($\sim10^9 M_{\rm sun}$) is suppressed by a factor of 
$\sim 10$ compared with the standard $\Lambda$CDM model. 
We then study the models with $k_{\rm cut} \simeq 51,\,410,\,820\,h\,{\rm Mpc}^{-1}$,
and confirm that the halo and the subhalo abundances and the radial distributions
of subhalos are indeed similar between the different WDM models and the Long-Lived CHAMP model.
The result suggests that the cut-off scale $k_{\rm cut}$ not only characterises the linear
power spectra but also can be used to predict the non-linear 
clustering properties.
The radial distribution of subhalos in Milky Way-size halos is
consistent with the observed distribution for 
$k_{\rm cut} \sim 50-800\,h\, {\rm Mpc}^{-1}$; such models resolve the so-called 
``missing satellite problem".
\end{abstract}

\keywords{cosmology: theory - early universe - dark matter }

\section{Introduction}

The precise measurement of the cosmic microwave background (CMB) anisotropies 
established the standard $\Lambda$ + Cold Dark Matter 
($\Lambda$CDM) cosmology \citep{2011ApJS..192...18K}. 
Observations of the large-scale structure of the Universe, 
such as the galaxy power spectra from the Sloan Digital Sky Survey (SDSS) also 
confirmed its success in predicting the large scale structures of the Universe\,
(e.g. \citet{2004PhRvD..69j3501T, 2010MNRAS.404...60R, 2010MNRAS.401.2148P}). 

The validity of the $\Lambda$CDM model on the galactic and the subgalactic scales 
has long been caught up in debate.
\citet{1999ApJ...524L..19M} argue that the number of dark matter subhalos 
is $10-100$ times larger than the number of satellites 
observed around the 
Milky Way\,\citep{2010AdAst2010E...8K}. 
The so-called ``missing satellite problem" has been revisited in a somewhat 
quantitative 
context \citep{2011MNRAS.415L..40B, 2012MNRAS.420.2318L, 2012MNRAS.422.1203B}. 
For example, \citet{2011MNRAS.415L..40B} argue that, in the $\Lambda$CDM model, 
$\sim10$ most massive subhalos in a galactic halo are too concentrated to be 
consistent with the kinematic data for the bright Milky Way satellites. 
Also, observations of the rotation velocities of 
galaxies using the $21$\,cm line by \,\citet{2011ApJ...739...38P}
show that the abundance of galaxies with observed velocity 
width $w=50\,{\rm km\,s^{-1}}$ is $\sim 8$ times lower than 
predicted in the $\Lambda$CDM model. 

It is often suggested that WDM models resolve 
the apparent problems on subgalactic scales\,\citep{2001ApJ...556...93B}.
WDM particles have non-negligible velocity dispersions, 
which act as an effective ``pressure'' 
of the WDM fluid. Essentially, the subgalactic-scale density fluctuations 
are suppressed.
The resultant matter power spectrum is quickly reduced around the cut-off scale 
that is determined by the velocity dispersion. 
Motivated by the recent interest in this problem, several authors study 
the structure formation in 
WDM models\,\citep{2011arXiv1109.6291D, 2012MNRAS.424..684S, 2012MNRAS.421.2384M}.

Constraints on WDM models can be obtained from astronomical observations.
Observations of Lyman-$\alpha$ forests are often used for the purpose\,\citep{2005PhRvD..71f3534V, 2009JCAP...05..012B}. 
Absorption features in quasar spectra reflect the number density of neutral 
hydrogen, from which we can estimate
the matter power spectrum along the line of sight, even at large wavenumbers $k \sim 10\,h\,{\rm Mpc}^{-1}$.
WDM models have also interesting implications for the cosmic reionization\,
\citep{2001ApJ...558..482B, 2003ApJ...591L...1Y, 2007Sci...317.1527G}. 
The formation of the first objects, and hence the production of ionizing photons, 
are delayed in WDM models.
On the other hand, WDM models could help the completion of the cosmic reionization. 
\citet{2012ApJ...747..127Y} suggest that the reduced number of subhalos 
in WDM models makes the recombination 
of ionized hydrogens inefficient and results in earlier completion of the cosmic 
reionization.
It is clearly important to study
the clustering properties in WDM models in both linear and non-linear 
evolution regimes.

There are also renewed interest in particle physics. 
Several candidates for WDM are suggested in particle physics models beyond the 
Standard Model, 
such as light gravitinos\,\citep{1997MPLA...12.1275K}, 
sterile neutrinos\,(see \citet{2009PhR...481....1K} 
for a review and references) and superWIMPs\,\citep{2005PhRvL..95r1301C}.
It is important to notice that WDM particles can be produced via different mechanisms. 
Nevertheless, the above constraints from astronomical observations are focused on
a single quantity, e.g., the mass of WDM particle in a specific model.
It is unclear if such constraints can be applied to WDM models 
with different production
mechanisms. Detailed comparisons of a wide class of models are clearly needed.

In this paper, we also consider a Long-Lived CHAMP model.
Throughout this paper, we assume that CHAMPs have an elementary charge, 
either positive or negative.
CHAMPs are generally realized in models beyond the Standard Model of particle 
physics.
One such example is a slepton, a superpartner of leptons in supersymmetric models.
Sleptons as the lightest supersymmetry particles (LSPs) are stable when R-parity 
is conserved.
The abundance of such stable CHAMPs, however, is severely constrained by the searches in deep 
sea water\,(see \citet{Beringer:1900zz} for a review and references).
CHAMPs can also be unstable; a CHAMP decay into neutral dark matter and other decay products 
including at least one charged particle. 
For example, the stau can be the next lightest supersymmetric particle (NLSP) when the gravitino 
is the LSP\,\citep{2006JCAP...11..007B}.
It is well-known that CHAMPs could affect the big bang nucleosynthesis (BBN) reaction rates 
and thus change the abundance of light elements\,\citep{2007PhRvL..98w1301P, 
2007PhRvD..76f3507K, 2006PhRvD..74j3004K, 2006JCAP...11..014C, 2007AIPC..903..595S, 
2007PhLB..650..268H, 2007PhLB..649..436K, 2008PhRvD..77f3524J, 2008JCAP...03..008J, 2011PhRvD..84c5008J}.
Several authors\,\citep{2004PhRvL..92q1302S, 2010PhLB..682..337K} suggest the possibility 
that CHAMPs with a lifetime about $1\,{\rm yr}$ can act effectively as WDM
through acoustic oscillations in the thermal background.
We study the effect of the oscillations on the matter power spectrum.

We calculate the linear evolution of the matter density fluctuations for the three WDM models 
and the Long-Lived CHAMP model.
We show that the comoving Jeans scale at the matter-radiation equality characterises the 
linear matter power spectra in the three WDM models well.
We use the obtained linear matter power spectra as initial conditions of $N$-body simulations 
to follow the non-linear evolution of the matter distribution.
We compare the halo mass functions, the subhalo mass functions, and the radial distributions 
of subhalos 
in Milky Way-size halos to discuss the clustering properties in the WDM models and in the Long-Lived CHAMP model.
We show that these statistics are similar when the cut-off scale is kept the same.
We find that the WDM models and the Long-Lived CHAMP model with the characteristic cut-off scale 
$k_{\rm cut} \sim 50-800\,h\, {\rm Mpc}^{-1}$ resolve the so-called 
``missing satellite problem".

The rest of this paper is organised as follows.
In Sec.\,\ref{sec:models}, we summarize three WDM models and a Long-Lived CHAMP model we consider.
Then, we introduce the common cut-off scale $k_{\rm cut}$ which characterises the linear matter power spectra in these models.
In Sec.\,\ref{sec:numericalsimulations}, after describing the details of $N$-body simulations, we show simulation results and discuss their implications.
Specifically, we mention the similarity of these models with the same cut-off and the possibility that CHAMPs 
behave like WDMs and resolve the ``missing satellite problem". 
Finally, in Sec.\,\ref{sec:summary}, concluding remarks are given.

Throughout this paper, we take the cosmological parameters 
that are given in \citet{2011ApJS..192...18K} 
as the {\it WMAP}+BAO+$H_{0}$ Mean; $100 \Omega_{b} h^{2}=2.255$, 
$\Omega_{\rm CDM}h^{2}=0.1126$, $\Omega_{\Lambda}=0.725$, 
$n_{s}=0.968$, $\tau=0.088$ and $\Delta^{2}_{R}(k_{0})=2.430\times10^{-9}$, 
while we replace the energy density of CDM $\Omega_{\rm CDM}h^{2}$
by the energy density of WDM $\Omega_{\rm WDM}h^{2}$ for the WDM models 
and by the energy density of neutral dark matter produced by the
CHAMP decay for the Long-Lived CHAMP model.

\section{WDM models and Long-Lived CHAMP model}
\label{sec:models}
In this section, we summarize three WDM models and a Long-Lived CHAMP model 
we consider in this paper.
We describe production mechanisms of WDM particles in each model and show 
the exact shapes of the velocity distribution.
In the following subsections, we focus on three WDM models to 
specify our discussion, although our results can be applied to 
any WDM models with the same shape of the velocity distribution. 
Then, we introduce the Jeans scale at the matter-radiation equality.
The matter power spectra in the three WDM models with the same Jeans scale 
at the matter-radiation equality are very similar.
Their initial velocity distributions affect the damping tail of the matter power spectra.
We also describe the evolution of the linear matter density fluctuations 
in a Long-Lived CHAMP model.
The matter power spectrum is truncated around the horizon scale at the time when CHAMPs
decay. Interestingly, the resulting power spectrum appears similar to those 
in WDM models.

\subsection{Thermal WDM}
\label{subsec:ThermalWDM}
In this type of models, fermionic WDM particles are produced in the thermal background.
They are decoupled from the thermal background as the Universe expands and cools.
At the time of the decoupling, their momentum obeys the thermal distribution, 
that is, the Fermi-Dirac distribution.
We consider the generalized Fermi-Dirac distribution,
\begin{eqnarray}
f(p) = \frac{\beta}{e^{p/T_{\rm WDM}}+1}\,.
\label{eq:fermi}
\end{eqnarray}
Here and in the following, $p$ denotes the comoving momentum of WDM particles, and
$T_{\rm WDM}$ is the effective temperature that characterises the comoving momentum of WDM particles.
In the case of the light gravitino\,\citep{1996PhRvD..53.2658D} and 
the thermally produced sterile neutrino\,\citep{1982PhRvD..25..213O}, 
$T_{\rm WDM}$ relates to the temperature of the left-handed neutrino $T_{\nu}$ 
through the conservation of the entropy, $T_{\rm WDM}=\left( \frac{43/4}{g_{\rm dec}} \right)^{1/3}T_{\nu}$ 
where $g_{\rm dec}$ is the effective number of the massless degrees of freedom at the decoupling from the thermal background. 
Note that $\beta$ determines the overall normalization of the momentum distribution and $\beta=1$ 
in the case of the gravitino and the thermally produced sterile neutrino.
\citet{1994PhRvL..72...17D} propose the sterile neutrino dark matter produced via active-sterile neutrino oscillations.
In this case, the active neutrinos in the thermal background turn into the sterile neutrino via the coherent forward scattering\,\citep{1992PhRvL..68.3137C}.
The resultant momentum distribution of the sterile neutrino is given by the generalized Fermi-Dirac distribution 
(see Eq.\,(\ref{eq:fermi})) with $T_{\rm WDM} \simeq T_{\nu}$ and $\beta \propto \theta^2_{\rm m} M$ where $\theta_{\rm m}$ is the active-sterile mixing 
angle and $M$ is the mass of the sterile neutrino.

\subsection{WDM produced by the thermal boson decay}
There are models in which the Majorana mass of the sterile neutrino arises from the Yukawa coupling $Y$ 
with a singlet boson\,\citep{2006PhLB..639..414S, 2008PhRvD..77f5014P}. 
In these models, the singlet boson couples to the Standard Model Higgs boson through an extension of the Standard Models Higgs sector.
The singlet Higgs boson has a vacuum expectation value (VEV) of the order of the electroweak scale when the electroweak symmetry breaks down.
When the sterile neutrino is assumed to be WDM with a mass of an order of keV, the Yukawa coupling should be very small $Y \sim O(10^{-8})$.
This small Yukawa coupling makes the singlet boson decay to the two sterile neutrinos 
when the singlet boson is relativistic and is in equilibrium with the thermal background.
Here, it should be noted that the sterile neutrino model is one specific example.
In WDM models, where relativistic bosonic particles in equilibrium with the thermal background decay into
fermionic WDM particles through the Yukawa interaction, 
WDM particles have the same resultant momentum distribution (see Eq.\,(\ref{eq:fB}) below).
The resultant momentum distribution is obtained by solving the Boltzmann equation\,\citep{2008PhRvD..78j3505B},
\begin{eqnarray}
f(p) = \frac{\beta}{(p/T_{\rm WDM})^{1/2}}g_{5/2}(p/T_{\rm WDM})
\label{eq:fB}
\end{eqnarray}　
where
\begin{eqnarray}
g_{\nu}(x)=\sum_{n=1}^{\infty}\frac{e^{-nx}}{n^{\nu}}\,.
\end{eqnarray}
Here, we have ignored the low momentum cut-off 
that ensures the Pauli blocking, while it does not change our results.
The effective temperature is given by $T_{\rm WDM}=\left( \frac{43/4}{g_{\rm pro}} \right)^{1/3}T_{\nu}$ with the effective number of massless 
degrees of freedom at the production of the sterile neutrino $g_{\rm pro} \sim 100$.
The normalization factor $\beta$ is determined by the Yukawa coupling $Y$ and the mass of the singlet Higgs boson $M$, $\beta \propto Y^2M^{-1}$.
The velocity distribution have an enhancement $f_{\rm B}\propto p^{-1/2}$ at the low momentum $p/T_{\rm WDM}\ll1$, since the sterile neutrinos with lower momenta are produced by the less boosted singlet boson, the decay rate of which is larger due to the absence of the time dilation.
This enhancement indicates the ``colder'' (than the thermal WDM) property of the sterile neutrino dark matter produced by the decay of the singlet heavy boson.

\subsection{WDM produced by the non-relativistic particle decay}
In this type of models, a non-relativistic heavy particle decays into two particles, one or both of which become WDM. 
Supersymmetric theories realize this type of scenarios e.g. when the LSP is the gravitino and the NLSP is a neutralino.
The relic abundance of the NLSP neutralino is determined at the time of chemical decoupling by the standard 
argument\,\citep{1991NuPhB.360..145G, 1991PhRvD..43.3191G}.
Eventually, the non-relativistic neutralinos decay into LSP gravitinos that become WDM.
The particles produced by the decay of the moduli fields and of the inflaton fields are another candidates of this 
type of WDM\,\citep{2001PhRvL..86..954L, 2001PhLB..505..169H, 2006PhRvD..74d3519K, 2006PhRvL..96u1301E, 2008PhLB..660..100T}.
When we assume the heavy particle decays in the radiation dominated era, 
the momentum distribution of the decay products is given by\,\citep{2005PhRvD..72f3510K, 2007PhRvD..75f1303S, 2011JCAP...09..025A},
\begin{eqnarray}
f(p) =\frac{\beta}{(p/T_{\rm WDM})}\exp(-p^2/T^2_{\rm WDM})\, ,
\end{eqnarray}
where $T_{\rm WDM}$ is given by $T_{\rm WDM}=P_{\rm cm} a(t_{d})/a(t_{0})$ with the physical canter-of-mass momentum $P_{\rm cm}$, 
the scale factor $a(t)$ at the decay time $t_{d}$ and at the present time $t_{0}$.
We have defined $t_{d}$ as $H(t=t_{d})=1/2\tau$ where $H(t)$ is the Hubble parameter and $\tau$ is the lifetime of the heavy particle.

\subsection{Jeans scale at the matter-radiation equality}
Now, we introduce two quantities to characterise the property of WDM.
One is the present energy density of WDM, $\Omega_{\rm WDM} \equiv \frac{\rho_{\rm WDM}}{\rho_{\rm crit}}{\big |}_{t=t_{0}}$.
Throughout this paper, we assume WDM particles account for all of the dark matter, letting $\Omega_{\rm WDM}h^2=0.1126$.
Another important physical scale is the comoving Jeans scale at the matter radiation equality $t_{\rm eq}$,
\begin{eqnarray}
k_{\rm J}=a\sqrt{\frac{4\pi G \rho_{\rm M}}{\sigma^2}}{\bigg |}_{t=t_{\rm eq}} \,
\label{eq:kjeans}
\end{eqnarray}
with the gravitational constant $G$.
Here, $\rho_{\rm M}$ is the matter density and $\sigma^2$ 
is the mean square of the velocity of the dark matter particles
(see Eq.\,(\ref{eq:vmoment}) below).
Dark matter particles with $k_{\rm J}\sim100-1000\,{\rm Mpc}^{-1}$ are usually called WDM 
and expected to resolve the ``missing satellite problem''.

We note that, in the present paper, we do not consider whether or not a particular set of 
$\Omega_{\rm WDM}$ and $k_{\rm J}$ is in a viable region of the respective model. 
One such example is the gravitino WDM, a representative of the Thermal WDM model (see subsection\,\ref{subsec:ThermalWDM}). 
This model has only two parameters, the effective number of the massless 
degrees of freedom at the decoupling $g_{\rm dec}$ and the gravitino mass $m_{3/2}$, to set $\Omega_{\rm WDM}$ and $k_{\rm J}$.
When we assume $k_{\rm J}\simeq30\,{\rm Mpc}^{-1}$, these two parameters are determined as $g_{\rm dec}\simeq1000$ and $m_{3/2}\simeq 1\,{\rm keV}$.	The effective number of the massless degrees of freedom at the decoupling of the gravitino is at most
$g_{\rm dec}\sim200$ in the Minimal Supersymmetric Standard Model (MSSM), and hence, another mechanism such as entropy production is needed
to explain the gravitino WDM\,\citep{2011JHEP...04..077I, 2012arXiv1204.3499I}.

\subsection{Linear matter power spectra and Normalized velocity distribution}
We follow the evolution of the primordial adiabatic fluctuations for the three WDM models 
by modifying suitably the public software, {\tt CAMB}\,\citep{2000ApJ...538..473L}.
We adopt the covariant multipole perturbation approach
for the massive neutrino\,\citep{1995ApJ...455....7M, 2002PhRvD..66b3531L}.
We replace the Fermi-Dirac distribution of the massive neutrino by the momentum 
distributions of the WDM models discussed above.
Our approach is valid when the WDM particles are kinematically decoupled at the cosmic time of interest.
The Jeans scale of interest is around $k_{\rm J} \sim O(100)\,\rm{Mpc}^{-1}$.
The primordial fluctuation of this wavenumber enters the horizon at $T \sim O(10)\,\rm{keV}$.
In a large class of WDM models, WDM particles are kinematically decoupled 
before the QCD phase transition, $T_{\rm QCD}\sim100\,\rm{MeV}$,
and thus our calculation is valid.
\footnote{There are variants of WDM models in which WDM particles are produced 
by the non-relativistic particle decay at late epochs. The matter power spectrum could be affected
if the parent particles decay around the matter-radiation equality.
In this case, the matter density during the radiation-dominated era 
is mostly contributed by the parent (cold) particles, rather than by the decay products,
and then the density fluctuations of the cold, neutral parent particles
can grow logarithmically. Note that the effect is more pronounced
if the parent particles are charged (see subsection\,\ref{subsec:champ}).}

\begin{figure}[tb]
 \begin{minipage}{.49\linewidth}
\begin{center}
  \includegraphics[width=\linewidth]{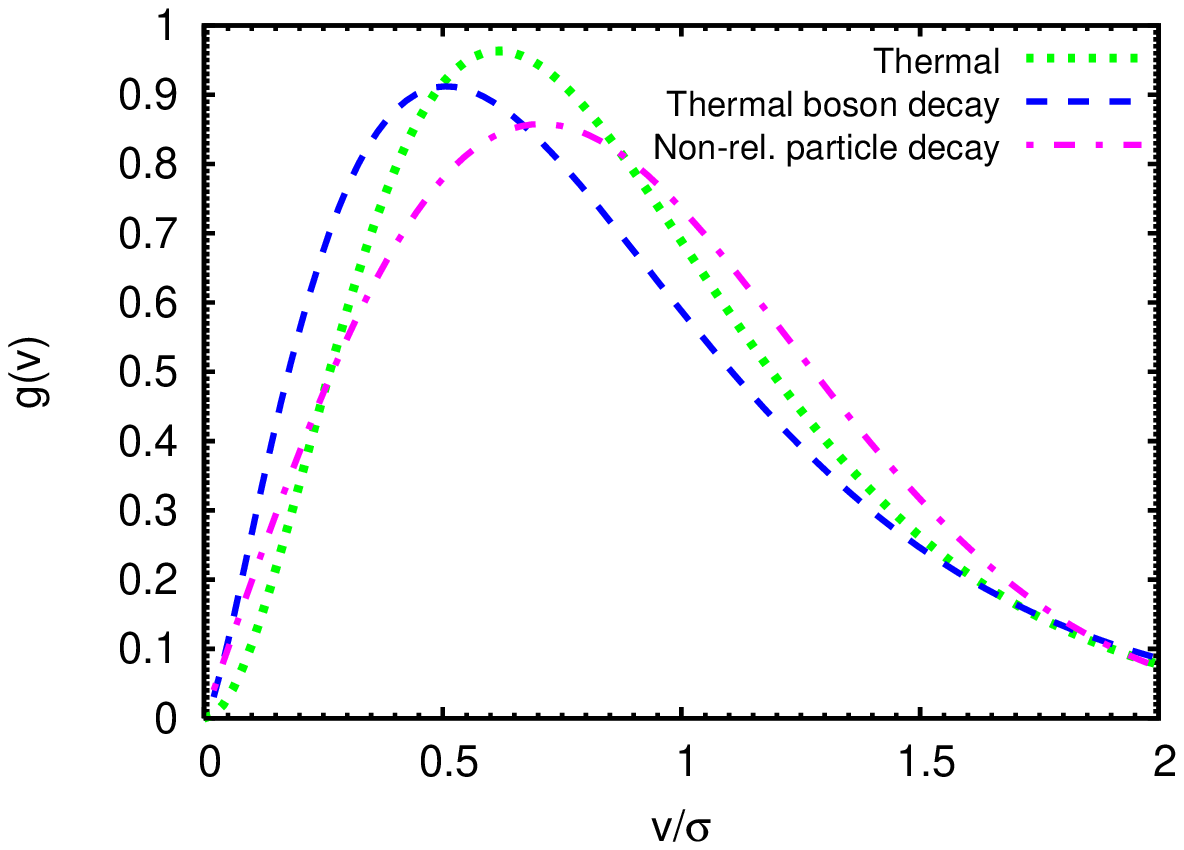}
  \end{center}
 \end{minipage}
 \begin{minipage}{.49\linewidth}
\begin{center}
  \includegraphics[width=\linewidth]{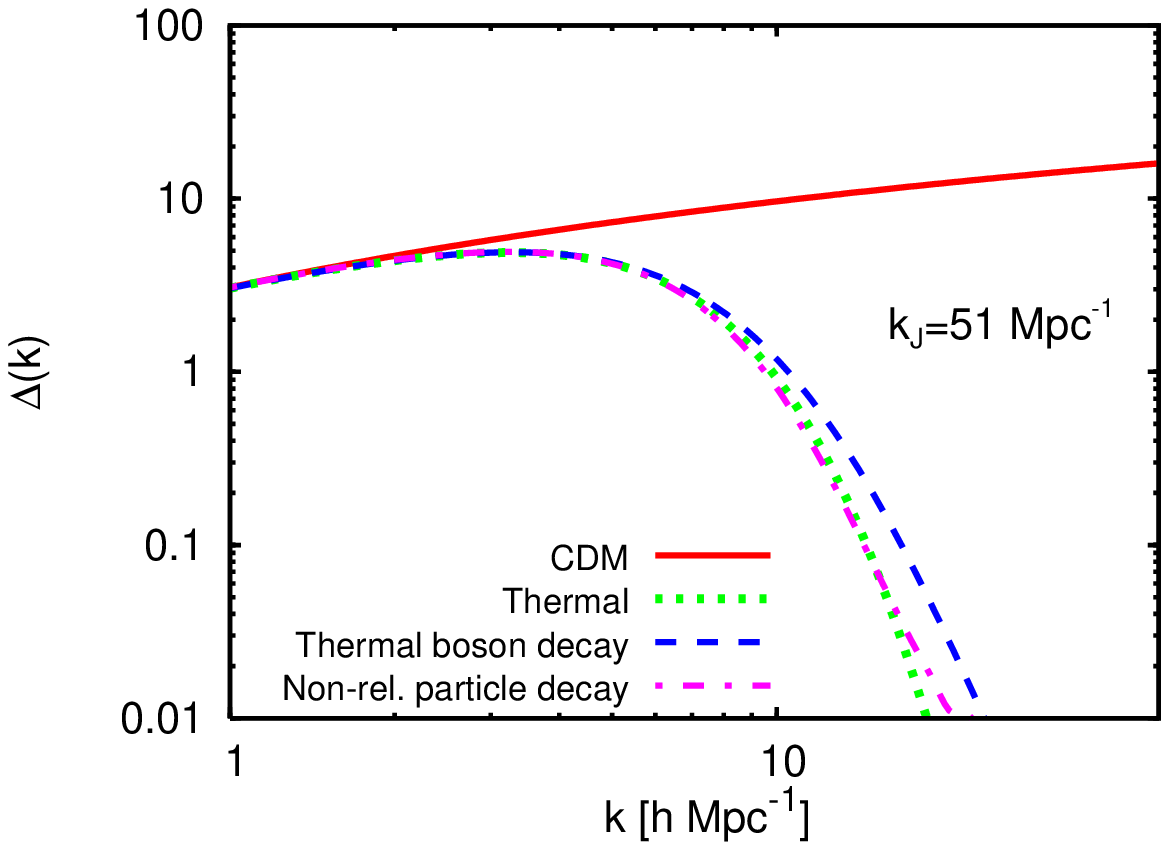}
  \end{center}
 \end{minipage}

\caption{
The normalized velocity distributions (left panel) and the dimensionless 
linear matter power spectra (right panel) for the standard CDM model 
and the three WDM models with $k_{\rm J}=51\,{\rm Mpc}^{-1}$.
}
\label{fig:power&velocity}
\end{figure}

For comparison, we calculate the normalized velocity distributions for the WDM models 
and the linear matter power spectra extrapolated to the present time $z=0$.
The results are shown in Fig.\,\ref{fig:power&velocity} for $k_{\rm J}=51\,{\rm Mpc}^{-1}$, 
which correspond to $m_{3/2} \simeq 2\,{\rm keV}$ for the thermally-produced gravitino WDM.
Here, we have defined the dimensionless matter power spectra as,
\begin{eqnarray}
\Delta(k) \equiv \frac{1}{2\pi^2}k^3P(k)
\end{eqnarray}
with the matter power spectra $P(k)$.

The velocity distribution $g(v)$ is normalized as follows:
\begin{eqnarray}
\int_{0}^{\infty} dv\,g(v) = 1\,, \\
\int_{0}^{\infty} dv\,v^2g(v) = \sigma^2\,
\label{eq:vmoment}
\end{eqnarray}
with the variance (second moment) of the velocity $\sigma^2$. 
Note that the first equation is normalized with respect to the present 
energy density of WDM, $\Omega_{\rm WDM}$,
whereas the second equation relates the velocity variance to
the comoving Jeans scale at the matter radiation equality $k_{\rm J}$ 
given by Eq.\,(\ref{eq:kjeans}).
One can then expect that the power spectra for the WDM models with the same
$\Omega_{\rm WDM}$ and $k_{\rm J}$ are very similar, as seen in Fig.\,\ref{fig:power&velocity}. 
There, we see differences between the WDM models only in the damping tail of the power spectra
at $k > 10\,h\,{\rm Mpc}^{-1}$.

\subsection{Long-Lived CHAMP and Cut-off scale}
\label{subsec:champ}
\citet{2004PhRvL..92q1302S} formulate the linearized evolution equations for fluctuations 
in a Long-Lived CHAMP model. They show that 
the subgalactic-scale matter density fluctuations 
are damped via a mechanism called ``acoustic damping''. 
The comoving horizon scale at which CHAMP decays determines the cut-off scale of the matter power spectrum, 
which is defined by\,\citep{2006PhLB..643..141H, 2010PhLB..682..337K}
\begin{eqnarray}
k_{\rm Ch} = aH|_{t=\tau_{\rm Ch}},
\end{eqnarray}
where $H$ is the Hubble parameter and $\tau_{\rm Ch}$ is the lifetime of CHAMP.
Smaller-scale density fluctuations with $k > k_{\rm Ch}$ enter the horizon before CHAMP decays and can not grow due to the acoustic oscillations of CHAMP in the thermal background.
On the other hand, larger-scale density fluctuations with $k < k_{\rm Ch}$ grow logarithmically even after entering the horizon due to the gravitational instability as the density fluctuations of CDM.
We assume CHAMPs decay in the radiation dominated era. 
Then the comoving horizon scale at $t = \tau_{\rm Ch}$ is evaluated as, 
\begin{eqnarray}
k_{\rm Ch} = 2.2\,{\rm Mpc}^{-1}\times \left( \frac{\tau_{\rm Ch}}{\rm yr} \right)^{-1/2} \left( \frac{g_{\rm Ch}}{3.363} \right)^{1/4}
\label{eq:kchamp}
\end{eqnarray}
where $g_{\rm Ch}$ is the effective number of massless degrees of freedom when CHAMP decays.

We need to consider three physical processes for the CHAMP model.
First, we describe the neutralization of CHAMP.
A positively charged particle may become neutral by forming a bound state with an electron $e$.
Its binding energy is, however, almost the same as the hydrogen, $E_{{\rm b}\,e} \simeq 13.6\,{\rm eV}$.
Hence, the positively charged particle keeps charged until its decay, since we assume CHAMP decays in the radiation dominated era.
A negatively charged particle may become neutral by forming a bound state with a proton $p$.
Its binding energy $E_{{\rm b}\,p} \simeq 25\,{\rm keV}$ is almost $m_{^{4}{\rm He}}/m_{p} \sim 2000$ times larger than $E_{{\rm b}\,e}$
and hence is expected to make the negatively charged particle neutral at $T\sim1\,{\rm keV}$.
However, Helium $^{4}{\rm He}$ is produced through BBN, 
with which a negatively charged particle may form a binding state.
Its binding energy $E_{{\rm b}\,^{4}{\rm He}} \simeq 337\,{\rm keV}$ is 
almost $(Z_{^{4}{\rm He}}/Z_{p})^{2} \times m_{^{4}{\rm He}}/m_{p} \simeq 16$ times larger than $E_{{\rm b}\,p}$. 
It should be noted that even a negatively charged particle bound with a proton is wrested by $^4{\rm He}$ through a 
charge-exchange reaction\,\citep{2009PThPh.121.1059K}.
Therefore, when the yield of CHAMP $Y_{\rm Ch}$ ($Y\equiv n/s$ with the number density $n$ and the entropy density $s$) 
is smaller than the yield of the Helium $Y_{^{4}{\rm He}}$, 
almost every negatively charged particle forms a binding state with a helium nuclei, 
which has one positive elementary charge\,\citep{2010PhLB..682..337K}.

Second, the decay products of CHAMP may lead to energy injection to the thermal background. 
The resulting injection energy density is constrained from the photodissociation 
of BBN\,\citep{2001PhRvD..63j3502K} and CMB y- and $\mu-$ parameters\,\citep{1993PhRvL..70.2661H}.
However, models with CHAMP with almost the same mass with neutral dark matter 
are not severely constrained by BBN nor by CMB. We focus on such an ``unconstrained" model.
Note that the small mass splitting ensures the relatively long lifetime of CHAMP and the ``coldness'' 
of neutral dark matter.

Finally, CHAMPs are tightly coupled with baryons before its decay.
\citet{2004PhRvL..92q1302S} assume $\theta_{\rm baryon}=\theta_{\rm Ch}$ where $\theta$ is the 
divergence of the fluid velocity.
This approximation is valid when the Coulomb scattering between baryons and CHAMPs is efficient,
i.e., CHAMPs and baryons are tightly coupled.
However, the constraints from the Catalyzed BBN essentially allow
only heavy CHAMP with $m_{\rm Ch} \gtrsim 10^6\,{\rm GeV}$ for $\tau_{\rm Ch} \gtrsim 10^3 $sec.
It is unclear if the Coulomb scattering between baryons and such heavy CHAMPs is efficient.
We have calculated the scattering efficiency 
and found that the tightly coupled approximation is indeed valid through the epoch of interest 
for $m_{\rm Ch} \lesssim 10^8\,{\rm GeV}$. The details of the calculations are found elsewhere\,\citep{inprep}.
In the following, we adopt the formulation given by \citet{2004PhRvL..92q1302S} using the tightly coupled 
approximation between baryons and CHAMPs.

\begin{figure}[tb]
\begin{center}
  \includegraphics[width=0.49\linewidth]{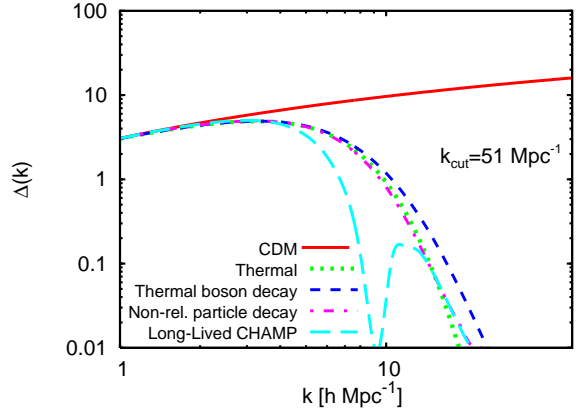}
\end{center}
\caption{\sl \small
We plot the dimensionless linear matter power spectrum in Long-Lived CHAMP model ($\tau \simeq 2.5 {\rm yr}$). 
We compare it with the same dimensionless linear matter power spectra in the WDM models as in Fig.\,\ref{fig:power&velocity}.
The oscillation around $k \sim 9\,h\,{\rm Mpc}^{-1}$ is the imprint of the ``acoustic damping''.
}
\label{fig:powerwithchamp}
\end{figure}

We modify {\tt CAMB}\,\citep{2000ApJ...538..473L} to follow the evolution of 
density fluctuations in the Long-Lived CHAMP model. The basic equations are given in 
\citet{2004PhRvL..92q1302S}. 
We obtain the power spectra for several $\tau_{\rm Ch}$s (see Eq.\,(\ref{eq:kchamp})). 
We find that the CHAMP matter power spectrum is very similar to the WDM models,
as seen in Fig.\,\ref{fig:powerwithchamp},
when $k_{\rm Ch}$ is set such that 
\begin{eqnarray}
\label{eq:kcut}
k_{\rm cut}\equiv k_{\rm J}\simeq 45\,k_{\rm Ch}\,.
\end{eqnarray}
Hereafter, we use $k_{\rm cut}$ defined in the above as a characteristic parameter
of the models we consider. 
The corresponding lifetime of CHAMP is $\tau_{\rm Ch}\simeq 2.5\,{\rm yr}$ in the figure.
The imprint of the CHAMP ``acoustic damping'' on the linear matter power spectra is clearly seen.
One can naively guess that structures in the Long-Lived CHAMP model would be similar to those
in the WDM models. 
It is important to study the non-linear growth of the matter distributions in the Long-Lived CHAMP model.
We use large cosmological $N$-body simulations to this end.

\section{Numerical simulations}
\label{sec:numericalsimulations}
Our simulation code is the parallel Tree-Particle Mesh code, {\tt GADGET-2}\,\citep{2005MNRAS.364.1105S}. 
We use $N=512^{3}$ particles in a comoving volume of $L=10\,h^{-1}\,{\rm Mpc}$ on a side. 
The mass of a simulation particle 
is $5.67\times10^{5}\,h^{-1}\,M_{\rm sun}$ and the gravitational softening length is $1\,h^{-1}\,{\rm kpc}$. 
We run a friends-of-friends (FoF) group finder\,\citep{1985ApJ...292..371D} to locate groups of galaxies. 
We also identify substructures (subhalos) in each FoF group using SUB-FIND algorithm developed by \citet{2001MNRAS.328..726S}.
We do not assign any thermal velocity to simulation particles because it can lead to formation of spurious objects\,\citep{2008ApJ...673..203C}.
We start our simulation from relatively low redshift $z=19$, at which the thermal motion of WDM is redshifted and negligible.
It should be noted that the heavy, neutral dark matter produced by the CHAMP decay is assumed to 
have negligible thermal velocities.

\begin{figure}[tb]
 \begin{minipage}{.33\linewidth}
 \begin{center}
 \includegraphics[width=\linewidth]{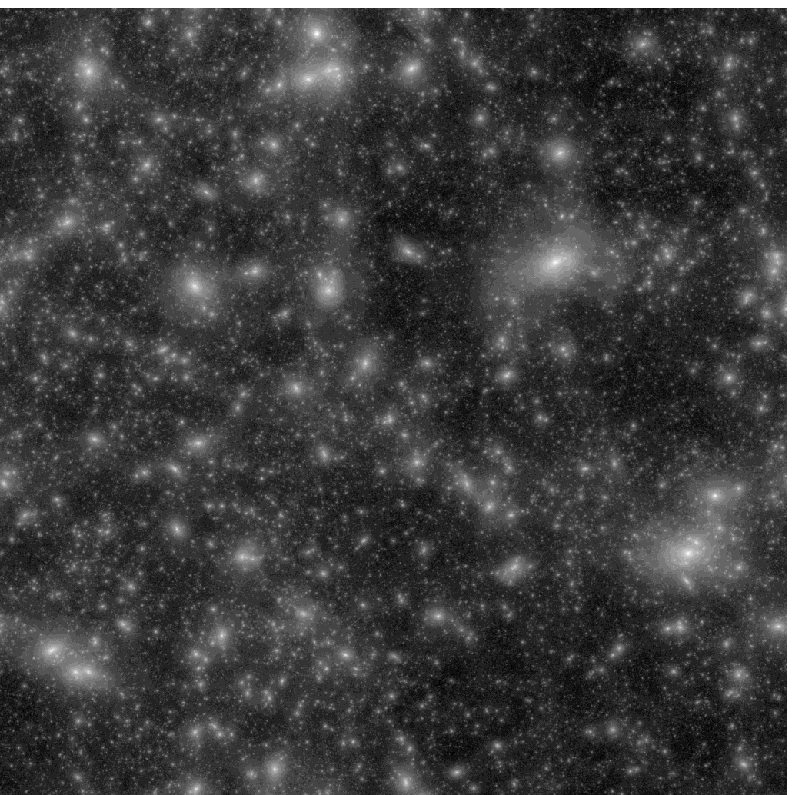}
 \end{center}
 \end{minipage}
 \begin{minipage}{.33\linewidth}
 \begin{center}
 \includegraphics[width=\linewidth]{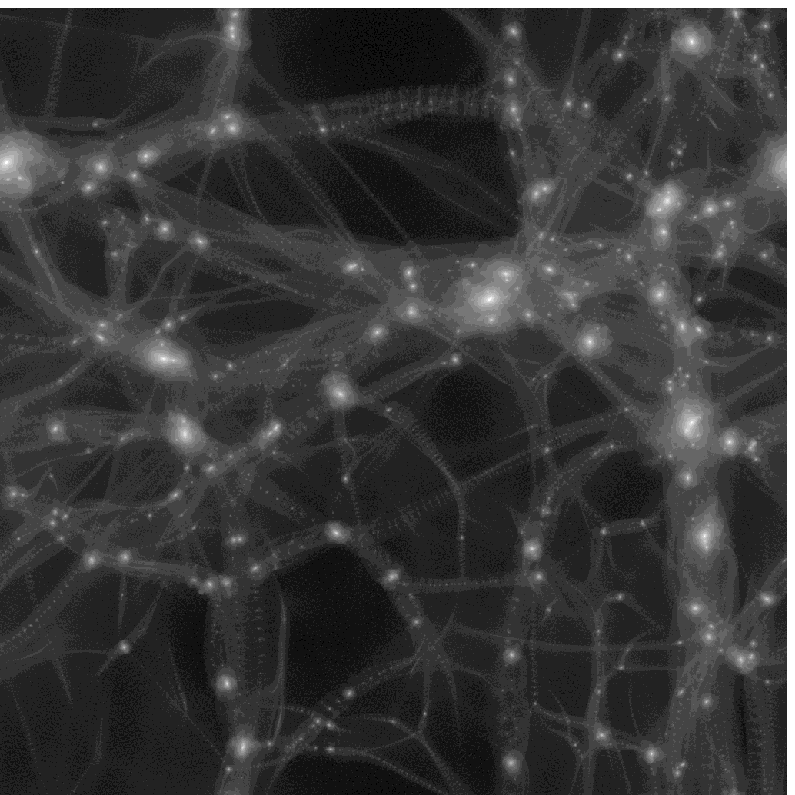}
 \end{center}
 \end{minipage}
 \begin{minipage}{.33\linewidth}
 \begin{center}
 \includegraphics[width=\linewidth]{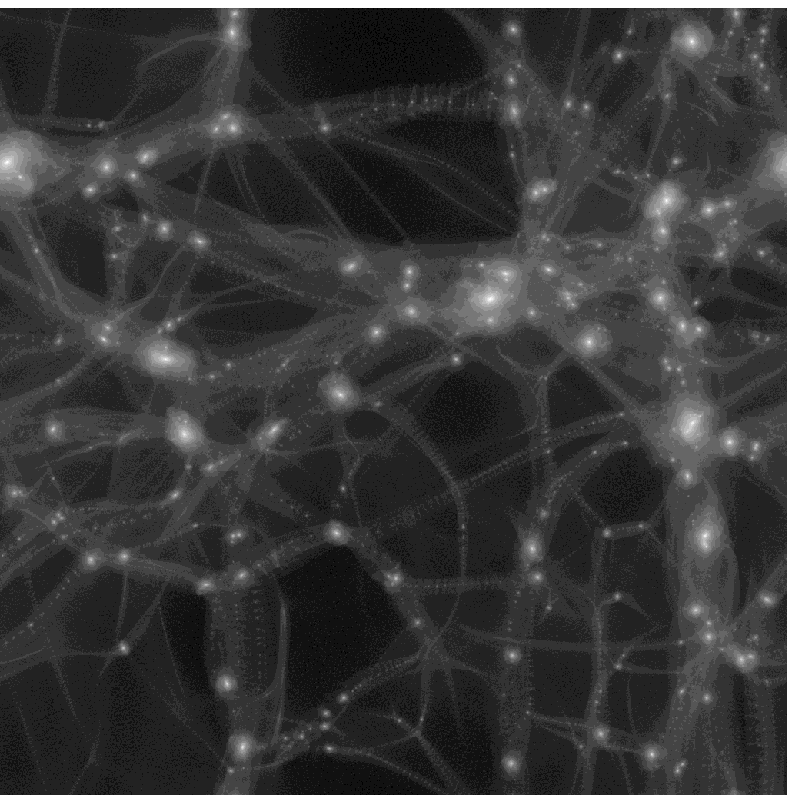}
 \end{center}
 \end{minipage}
 \caption{\sl \small
 The projected matter distribution in the CDM model (left panel), in the Thermal WDM model (middle panel) and in the Long-Lived CHAMP model (right panel).
 For the Thermal WDM model and for the Long-Lived CHAMP model, we take the same cut-off scale $k_{\rm cut}=51\,{\rm Mpc}^{-1}$ as in Fig.\,\ref{fig:powerwithchamp}.
One side of the plotted region is $L=10\,h^{-1}\,{\rm Mpc}$.
Brighter regions denote higher matter densities. 
}
\label{fig:projectedplot}
\end{figure}

In Fig.\,\ref{fig:projectedplot}, we plot the projected matter distribution 
in the CDM model (left panel), in the Thermal WDM model (see subsection\,\ref{subsec:ThermalWDM}) (middle panel) and in the Long-Lived CHAMP 
model (right panel).
For the Thermal WDM model and for the Long-Lived CHAMP model, we set the same cut-off scale 
$k_{\rm cut}=51\,{\rm Mpc}^{-1}$ as in Fig.\,\ref{fig:powerwithchamp}.
One side of the plotted region is $10\,h^{-1}\,{\rm Mpc}$.
Regions with high matter densities appear bright in the plot. 
We see that many small objects, i.e., halos and subhalos, have formed in the CDM model.
Contrastingly, in the Thermal WDM model and in the Long-Lived CHAMP model, the matter distribution is much smoother
and appears more filamentary. The abundance of small objects is much reduced.
Overall, the matter distributions in the Thermal WDM model and in the Long-Lived CHAMP model 
look similar.
Note that numerous small objects along the filaments in the Thermal WDM model 
and in the Long-Lived CHAMP model could be numerical artifacts;
this is a long-standing problem of hot/warm dark matter simulations due to 
discreteness effects\,\citep{2007MNRAS.380...93W, 2011PhRvD..83d3506P}.
Earlier studies propose a simple formula for the critical halo mass, 
\begin{eqnarray}
M_{\rm c} = 10.1 \times \rho_{\rm M}\,d_{\rm mean}\,k_{\rm peak}^{-2}\,.
\label{eq:cutoffmass}
\end{eqnarray}
below which the abundance of halos is unreliable.
Here, $d_{\rm mean} = L / N^{1/3}$ is the mean comoving distance between simulation particles 
and $k_{\rm peak}$ is the wavenumber at the maximum of the $\Delta (k)$.
We will discuss this point further in the following section. 

\begin{figure}[tb]
 \begin{minipage}{.49\linewidth}
 \begin{center}
 \includegraphics[width=\linewidth]{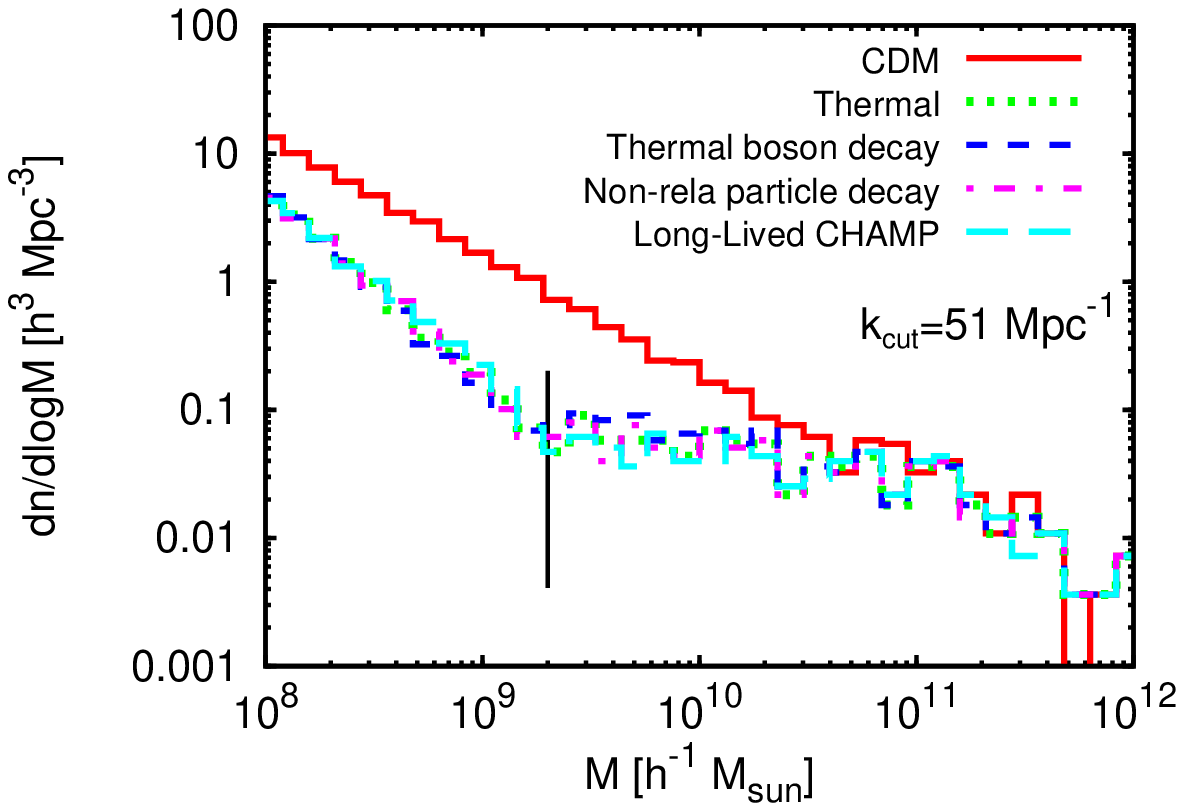}
 \end{center}
 \end{minipage}
 \begin{minipage}{.49\linewidth}
 \begin{center}
 \includegraphics[width=\linewidth]{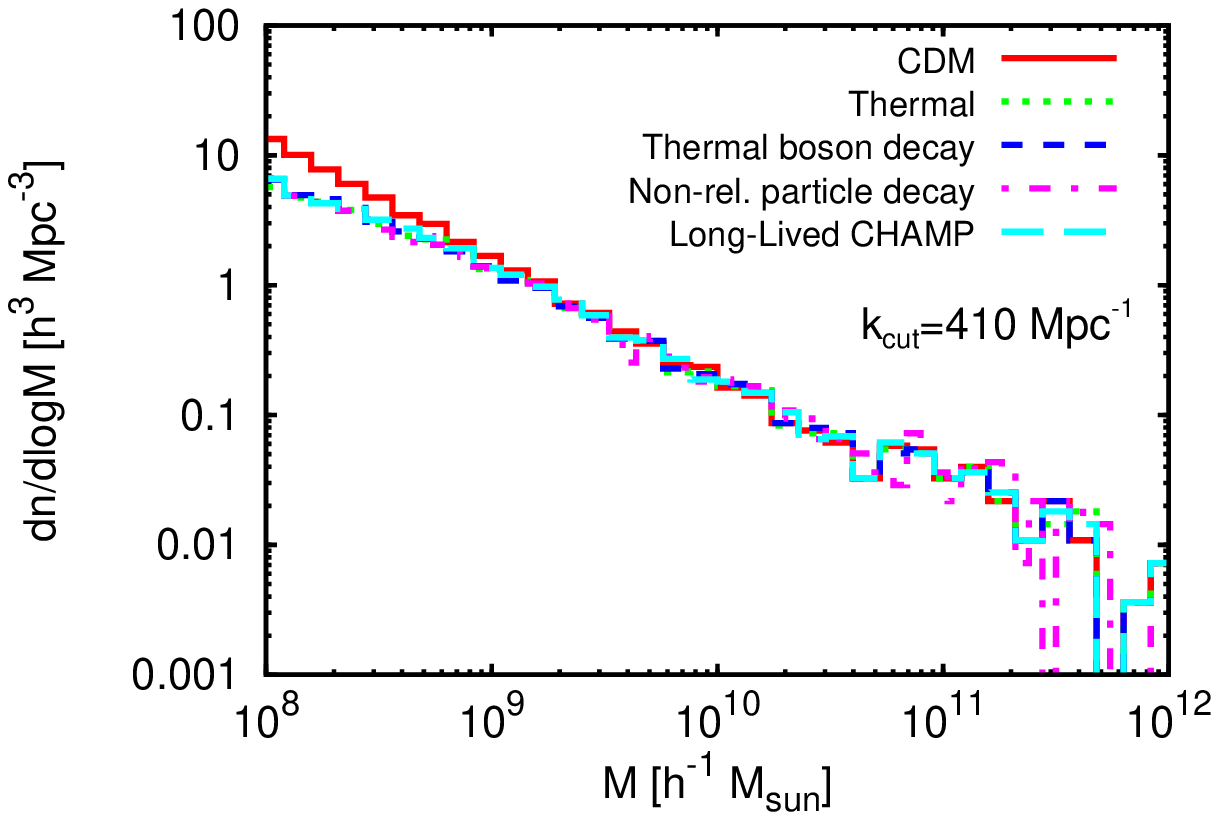}
 \end{center}
 \end{minipage}
 \caption{\sl \small 
 The halo mass functions in the CDM model, in the three WDM models and in the Long-Lived CHAMP model with $k_{\rm cut}=51\,{\rm Mpc}^{-1}$ (left panel) and with $k_{\rm cut}=410\,{\rm Mpc}^{-1}$ (right panel).
 The upturn at the halo mass $M\sim 2\times10^9\,h^{-1}\,M_{\rm sun}$ (vertical line) in the left panel may be owing to the artificial objects due to the discreteness effects.
}
\label{fig:massfunction}
\end{figure}

We compare the halo mass functions in the models we consider
in Fig.\,\ref{fig:massfunction}. 
The fiducial cut-off scale is $k_{\rm cut}=51\,{\rm Mpc}^{-1}$ (left panel)
as in Fig.\,\ref{fig:powerwithchamp},
but we also show the results for
$k_{\rm cut}=410\,{\rm Mpc}^{-1}$ (right panel).
The latter corresponds to $m_{3/2} \simeq 9.5\,{\rm keV}$ 
for the thermally-produced gravitino WDM (see subsection\,\ref{subsec:ThermalWDM})
and $\tau_{\rm Ch} \simeq 0.04\,{\rm yr}$ for the Long-Lived CHAMP model.
The halo mass $M$ corresponding to $k_{\rm cut}$ is given by 
\begin{eqnarray}
M = \frac{4\pi G \rho_{\rm M}}{3} \left( \frac{2\pi}{k_{\rm cut}} \right)^{3} \simeq 2\times10^{9}\,h^{-1}\,M_{\rm sun}\times \left( \frac{51\,{\rm Mpc}^{-1}}{k_{\rm cut}} \right)^{3}\,.
\end{eqnarray}

It is important to examine if the halo abundance is compromised by
the above-mentioned numerical artifacts.
For $k_{\rm cut}=51\,{\rm Mpc}^{-1}$, we see upturns in the mass functions 
at $M \sim 2\times10^9\,h^{-1}\,M_{\rm sun}$.
This is owing to peculiar discreteness effects in hot/warm dark matter simulations
\,\citep{2007MNRAS.380...93W, 2011PhRvD..83d3506P}.
The critical halo mass (see Eq.\,(\ref{eq:cutoffmass})) is 
$M_{\rm c} \simeq 2\times10^9\,h^{-1}\,M_{\rm sun}$ 
for our simulation parameters $L=10\,h^{-1}\,{\rm Mpc}$, $N=512^3$ and 
$k_{\rm peak} \simeq 3\,{\rm Mpc}^{-1}$ (see Fig.\,\ref{fig:powerwithchamp}). 
The estimated mass limit is indeed consistent with the upturn seen in the 
left panel of Fig.\,\ref{fig:massfunction}. 

Therefore, we conservatively restrict our discussion to halos with masses 
$M> M_{\rm c}=2\times10^9\,h^{-1}\,M_{\rm sun}$ for $k_{\rm cut}=51\,{\rm Mpc}^{-1}$.
The number of halos at $M \sim M_{\rm c}$ in the three WDM models and in the Long-Lived 
CHAMP model is $\sim 10$ times smaller than that in the CDM model.
Note the similarity of the halo abundances in the three 
WDM models and in the Long-Lived CHAMP model, as naively expected from the similarity 
in the linear matter power spectra. 

\begin{figure}[tb]
 \begin{minipage}{.49\linewidth}
 \begin{center}
 \includegraphics[width=\linewidth]{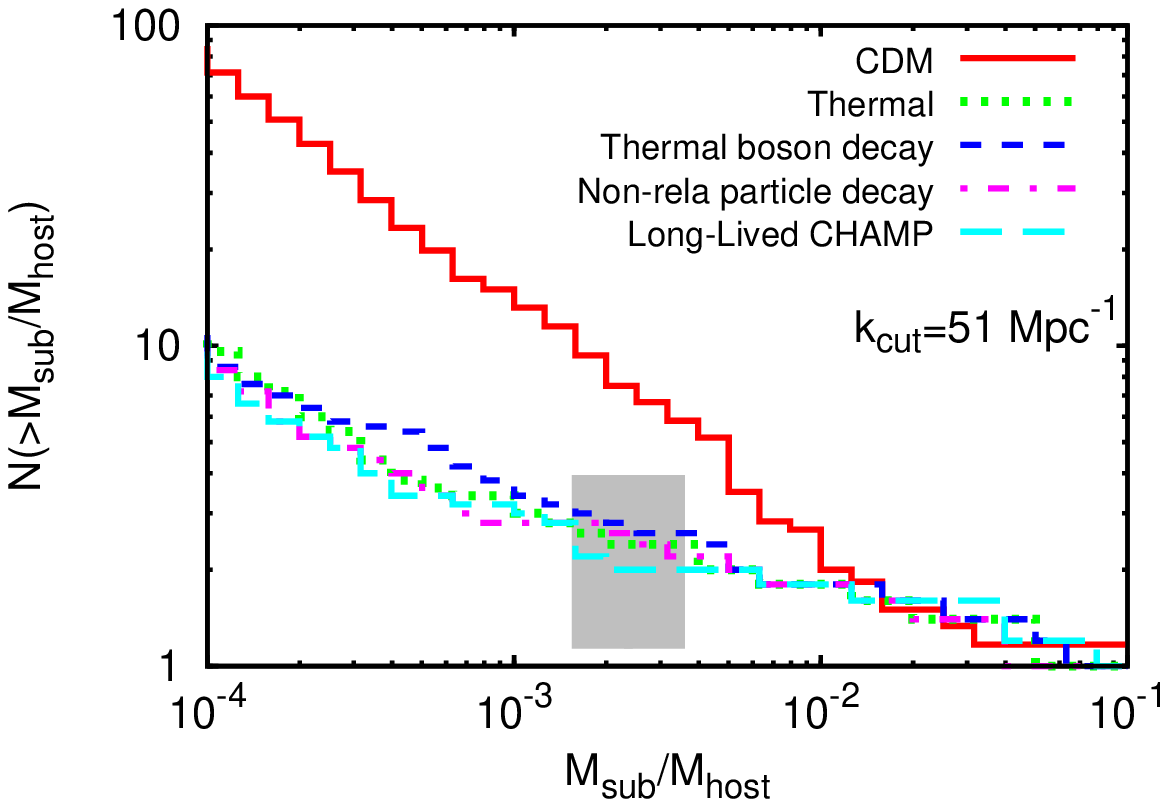}
 \end{center}
 \end{minipage}
 \begin{minipage}{.49\linewidth}
 \begin{center}
 \includegraphics[width=\linewidth]{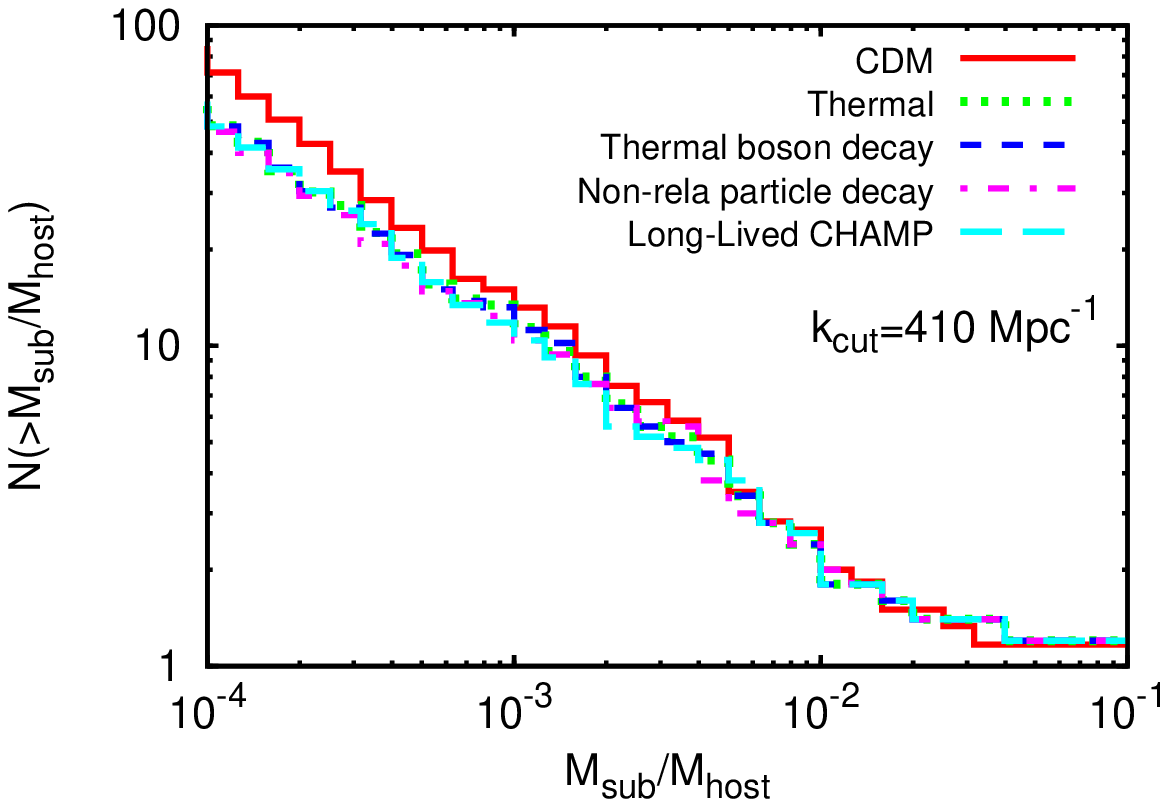}
 \end{center}
 \end{minipage}
 \caption{\sl \small 
 The cumulative subhalo mass functions averaged over Milky Way-size halos in the CDM model, in the three WDM models and in the Long-Lived CHAMP model with $k_{\rm cut}=51\,{\rm Mpc}^{-1}$ (left panel) and with $k_{\rm cut}=410\,{\rm Mpc}^{-1}$ (right panel).
 The shaded region corresponds to the mass of nonlinear objects at which the upturn (numerical artifacts) occurs in halo mass function (see the vertical line in the left panel of Fig.\,\ref{fig:massfunction}).
}
\label{fig:nomsubcum}
\end{figure}

\begin{figure}[tb]
 \begin{minipage}{.49\linewidth}
 \begin{center}
 \includegraphics[width=\linewidth]{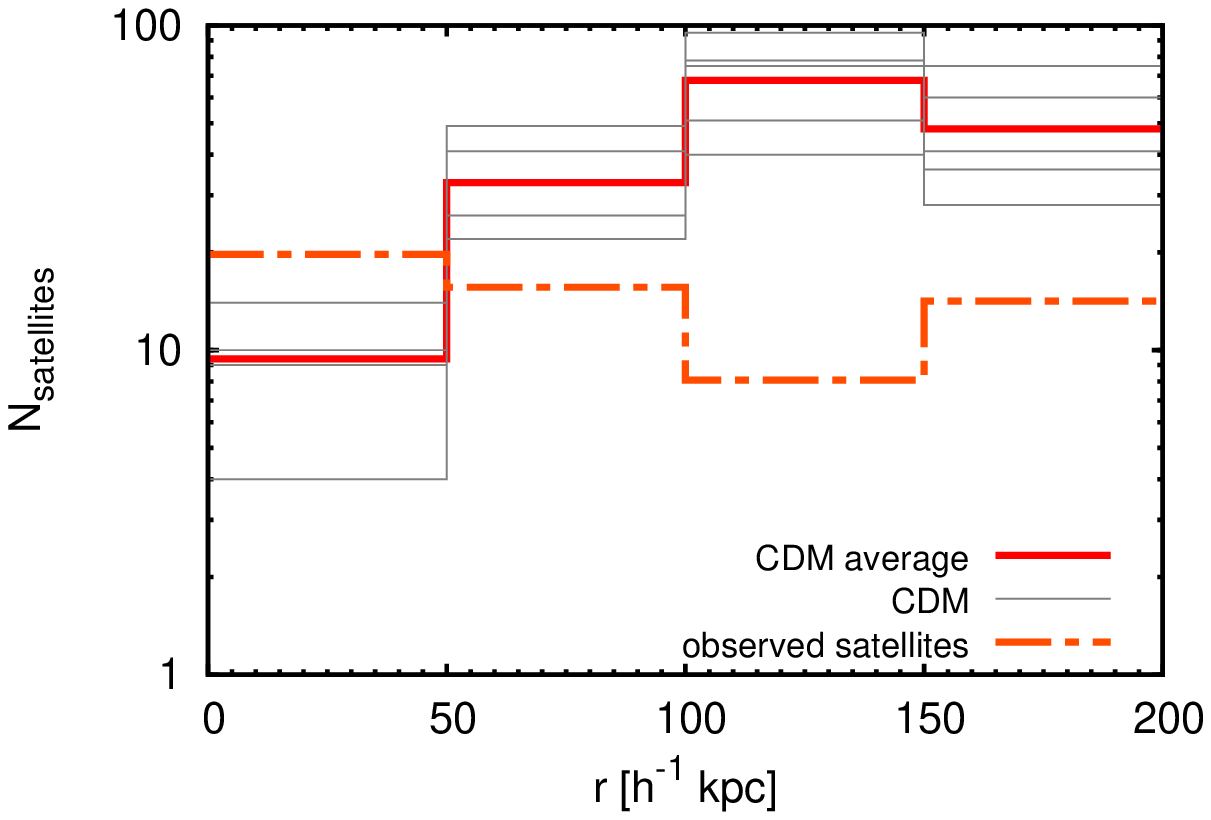}
 \end{center}
 \end{minipage}
 \begin{minipage}{.49\linewidth}
 \begin{center}
 \includegraphics[width=\linewidth]{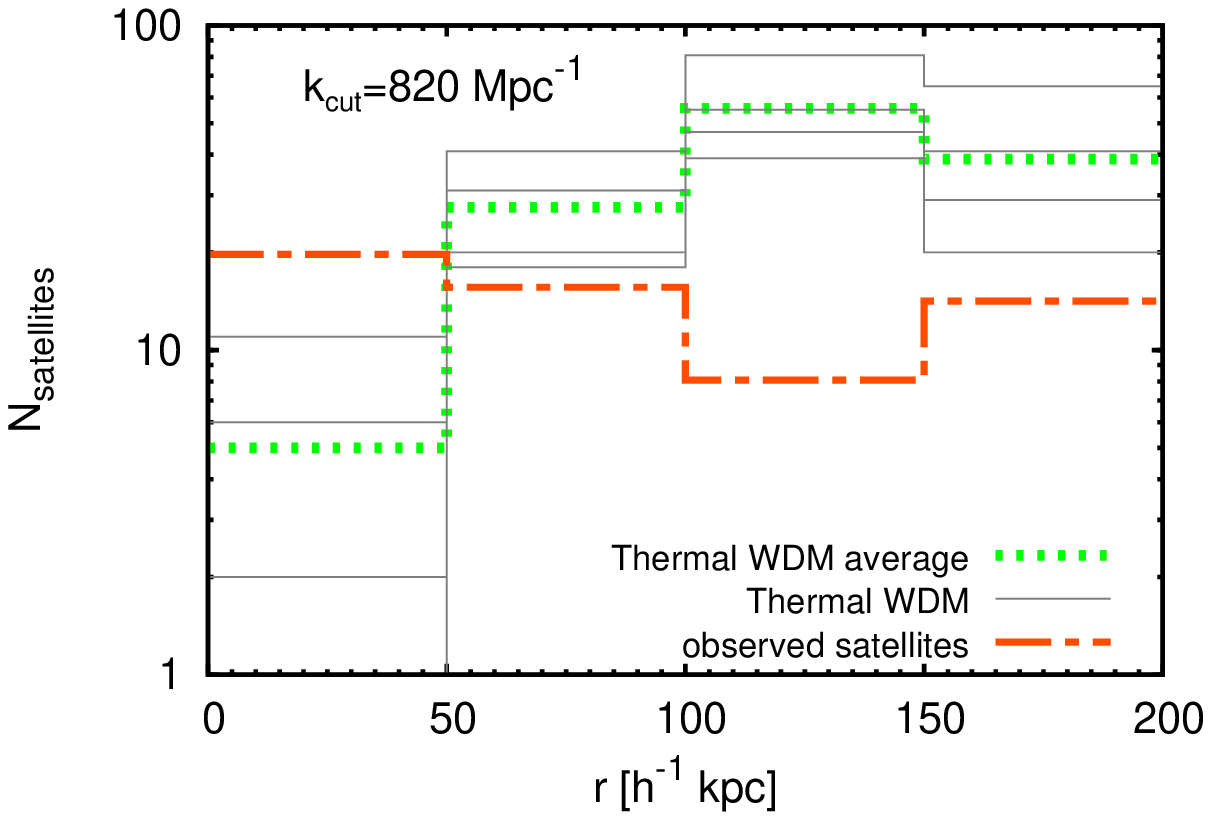}
 \end{center}
 \end{minipage}
 \caption{\sl \small
 The radial distributions of subhalos in Milky Way-size halos in the CDM model (right panel) and in the Thermal WDM models (left panel) with $k_{\rm cut} \simeq 810\,{\rm Mpc}^{-1}$.
 We divide the distance from the center of host halo in $50\,{\rm kpc}$ bins. Each thin line corresponds to the radial distribution in each Milky Way-size halo. 
 Thick lines represent the average over the Milky Way-size halos.
 For comparison, we also plot the radial distribution of the observed satellites listed in Table \ref{table:sdssdata}.
}
\label{fig:cdmradialdistirbution}
\end{figure}

\begin{table}[tb]
  \begin{center}
  \begin{tabular}{|c||c|c|c|c|} \hline
  distance from the center & $0-50$ & $50-100$ & $100-150$ & $150-200$\\ \hline
  number of satellites & $19.7$ & $15.62$ & $8.08$ & $14.16$ \\ \hline
  \end{tabular}
 \end{center}
 \caption{\sl \small
 The number of observed satellites in each $50\,{\rm kpc}$ from the center of our Milky Way. According to \citet{2011PhRvD..83d3506P}, we count observed satellites known before the SDSS as one and those found by the SDSS as 3.54 due to the limited sky coverage of SDSS.
  }
   \label{table:sdssdata}
\end{table}

In order to see if the WDM models and the Long-Lived CHAMP model 
resolve the ``missing satellite problem'', 
we select Milky Way-size halos with masses 
of $0.5\times10^{12}\,h^{-1}\,M_{\rm sun}< M_{\rm halo} <1.5\times10^{12}\,h^{-1}\,M_{\rm sun}$ 
in our simulations.
Although the halo mass of Milky Way itself 
is in debate (e.g.\,\citet{2008ApJ...684.1143X} and references therein),
we take a relatively lower mass among suggested values.
For a (slightly) small value of $M_{\rm halo}$, the relative mass ratio
$M_{\rm satellite}/M_{\rm halo}$ becomes larger. Then the apparent discrepancy
between the number of observed satellites and the simulated subhalo abundance
at a given mass scale becomes smaller. We thus choose the small Milky Way mass
as a ``conservative'' one.

We compare the cumulative subhalo mass functions averaged over
the Milky Way-size halos in our five models in Fig.\,\ref{fig:nomsubcum}. 
Note that we see again an upturn for $k_{\rm cut}=51\,{\rm Mpc}^{-1}$ around 
$M_{\rm sub}/M_{\rm host}\simeq 2\times10^{-3}$.
Above the mass scale, where we can measure the mass function robustly, 
the subhalo abundance is suppressed by a factor of $\sim 10$ 
in the models with $k_{\rm cut}=51\,{\rm Mpc}^{-1}$ (left panel) compared with the CDM model.
For the models with $k_{\rm cut}=410\,{\rm Mpc}^{-1}$ (right panel), the subhalo abundance is suppressed at most by a factor of $\sim 2$.

Let us now examine the radial distribution of the subhalos in our simulated
Milky Way-size halos.
We adopt the abundance of observed satellites in our Milky Way 
listed in Table\,\ref{table:sdssdata}.
We account for the sky coverage of SDSS as follows.
We count the number of the observed satellites in each $50\,{\rm kpc}$ bin such 
that each satellite known before the Sloan Digital Sky Survey (SDSS) is weighted 
as one whereas each satellite discovered by the SDSS is weighted as 3.54\,\citep{2011PhRvD..83d3506P}.

In Fig.\,\ref{fig:cdmradialdistirbution}, we compare the averaged radial 
distributions of the subhalos in the Milky Way-size halos.
The left panel shows the radial distribution in the CDM model
and the right panel shows the radial distribution in the Thermal WDM model with $k_{\rm cut} \simeq 810\,{\rm Mpc}^{-1}$, 
which corresponds to $m_{3/2} \simeq 16\,{\rm keV}$ for the thermally-produced gravitino WDM.
We include all the subhalos with $M > 2 \times 10^7\,h^{-1}\,M_{\rm sun}$.
We also show variation of the radial distribution by thin solid lines.
It should be noted that we set the gravitational softening length to be 
$1\,h^{-1}\,{\rm kpc}$. The subhalo count in the innermost bin
could have been affected by the spatial resolution.
We can see the CDM model predicts a larger number of subhalos by a factor of $2-10$ in each radial bin than observed.
Note also that the number of satellites roughly scales with the host halo mass.
Among the five Milky Way-size halos we selected, which have masses of 
$0.5 \times 10^{12}\,h^{-1}\,M_{\rm sun} <  M_{\rm halo}  < 1.5 \times 10^{12}\,h^{-1}\,M_{\rm sun}$,
the total number of subhalos 
differs by a factor of $\sim 3$.
By comparing the two panels in Fig.\,\ref{fig:cdmradialdistirbution}, we find that the radial 
distribution of subhalos in the Thermal WDM model with $k_{\rm cut} \simeq 810\,{\rm Mpc}^{-1}$ 
is similar to the one in CDM model and hence,  
the Thermal WDM model with $k_{\rm cut} \simeq 810\,{\rm Mpc}^{-1}$ does not seem to resolve the ``missing satellite problem''.

\begin{figure}[tb]
 \begin{center}
 \includegraphics[width=0.49\linewidth]{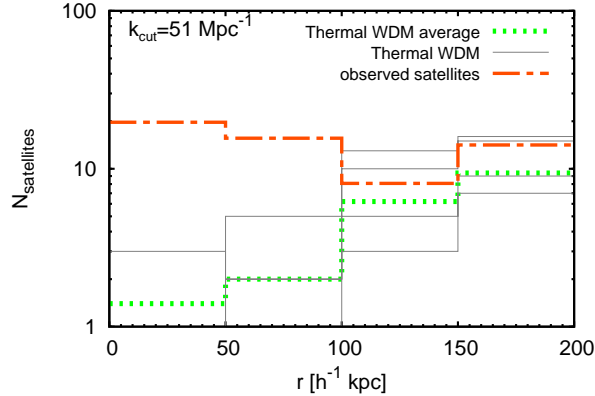}
 \end{center}
 \caption{\sl \small
 The radial distribution of subhalos in Milky Way-size halo in the Thermal WDM model with $k_{\rm cut} = 51\,{\rm Mpc}^{-1}$.
 This subhalos may include the artificial small objects due to the discreteness effects.
 }
\label{fig:twdmradialdistirbution}
\end{figure}

\begin{figure}[tb]
 \begin{minipage}{.49\linewidth}
 \begin{center}
 \includegraphics[width=\linewidth]{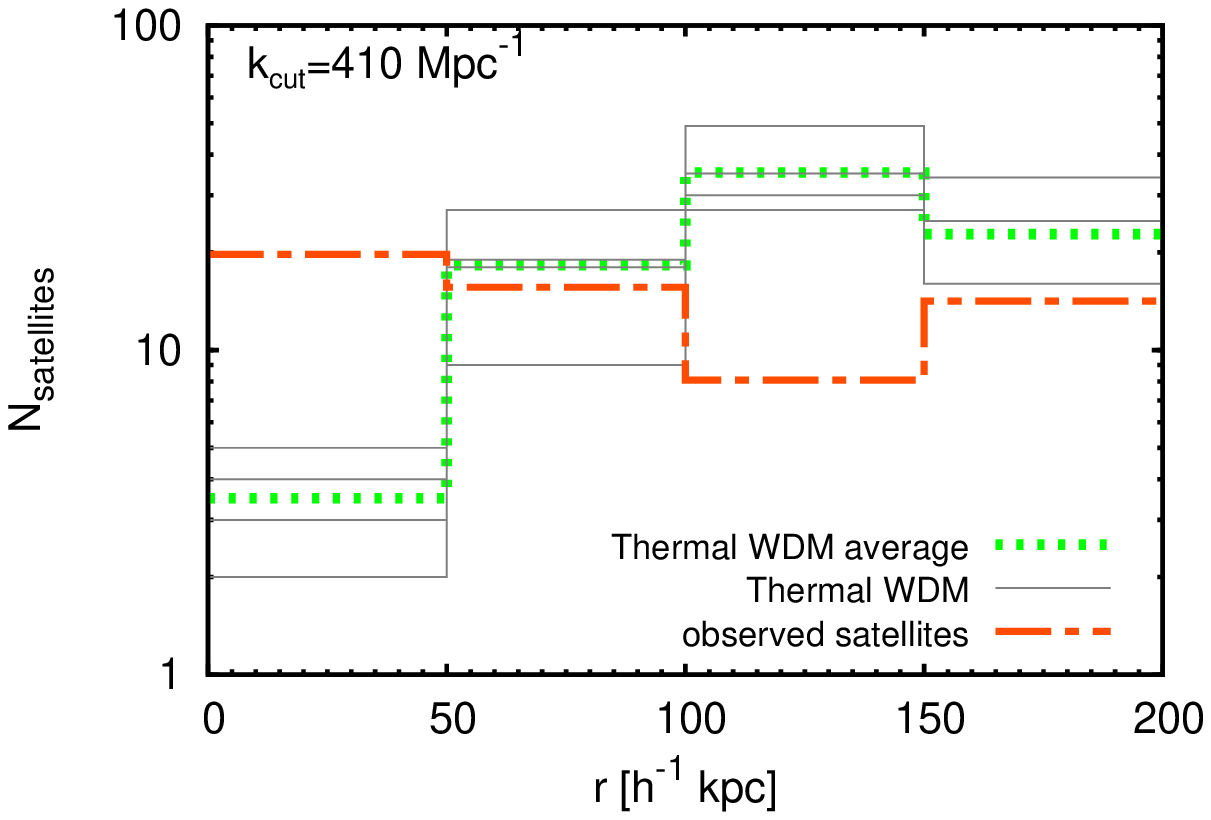}
 \end{center}
 \end{minipage}
 \begin{minipage}{.49\linewidth}
 \begin{center}
 \includegraphics[width=\linewidth]{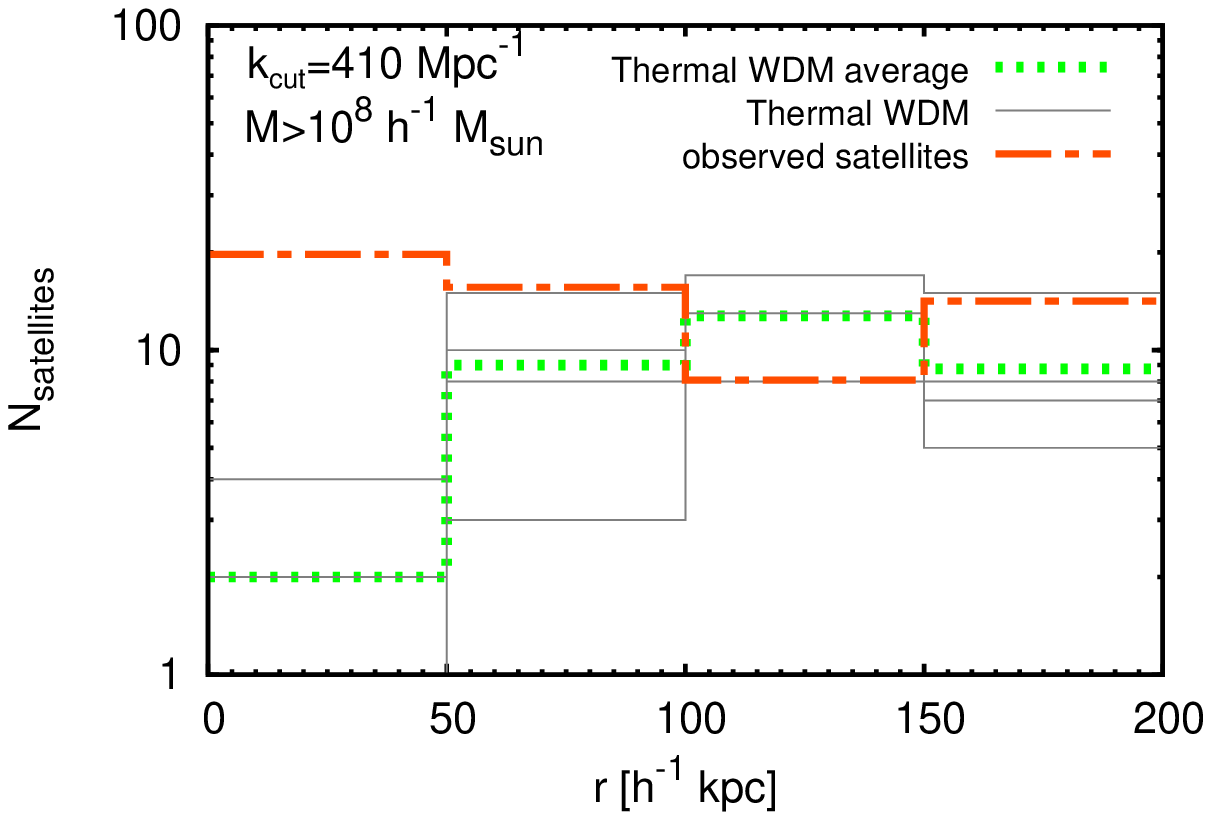}
 \end{center}
 \end{minipage}
 \caption{\sl \small
 The same plot as Fig.\,\ref{fig:twdmradialdistirbution} in the Thermal WDM model with $k_{\rm cut} = 410\,{\rm Mpc}^{-1}$. After discarding small subhalos ($M_{\rm sub} < M_{\rm c} \simeq 10^8\,h^{-1}\,M_{\rm sun}$), the subhalo abundance is reduced by a factor of $\sim 2$ (right panel).
}
\label{fig:twdm8radialdistirbution}
\end{figure}

We plot the subhalo radial distribution in the Thermal WDM model with $k_{\rm cut} \simeq 51\,{\rm Mpc}^{-1}$ 
in Fig.\,\ref{fig:twdmradialdistirbution}. For this model, where the suppression of the subgalactic-scale structure 
is most significant, a sizeable fraction of subhalos
have masses smaller than $M_{\rm c}$ (see Eq.\,(\ref{eq:cutoffmass})). 
Thus the number count in the radial distribution is likely unreliable. 
Note however that, even without discarding the small mass subhalos ($M_{\rm sub} < M_{\rm c}$),
the subhalo abundance is slightly smaller than the observed satellites.

We plot the radial distributions of subhalos with masses
$M>M_{\rm c} \simeq10^{8}\,h^{-1}\,M_{\rm sun}$ in the Thermal WDM model with $k_{\rm cut} \simeq 410\,{\rm Mpc}^{-1}$ in Fig.\,\ref{fig:twdm8radialdistirbution}.
For this model, discarding small subhalos ($M_{\rm sub} < M_{\rm c} \simeq 10^8\,h^{-1}\,M_{\rm sun}$)
reduces the subhalo abundance by a factor of $\sim 2$. Overall, 
the radial distribution of subhalos (after discarding) appears to reproduce the observed distribution.

\begin{figure}[tb]
 \begin{minipage}{.49\linewidth}
 \begin{center}
 \includegraphics[width=\linewidth]{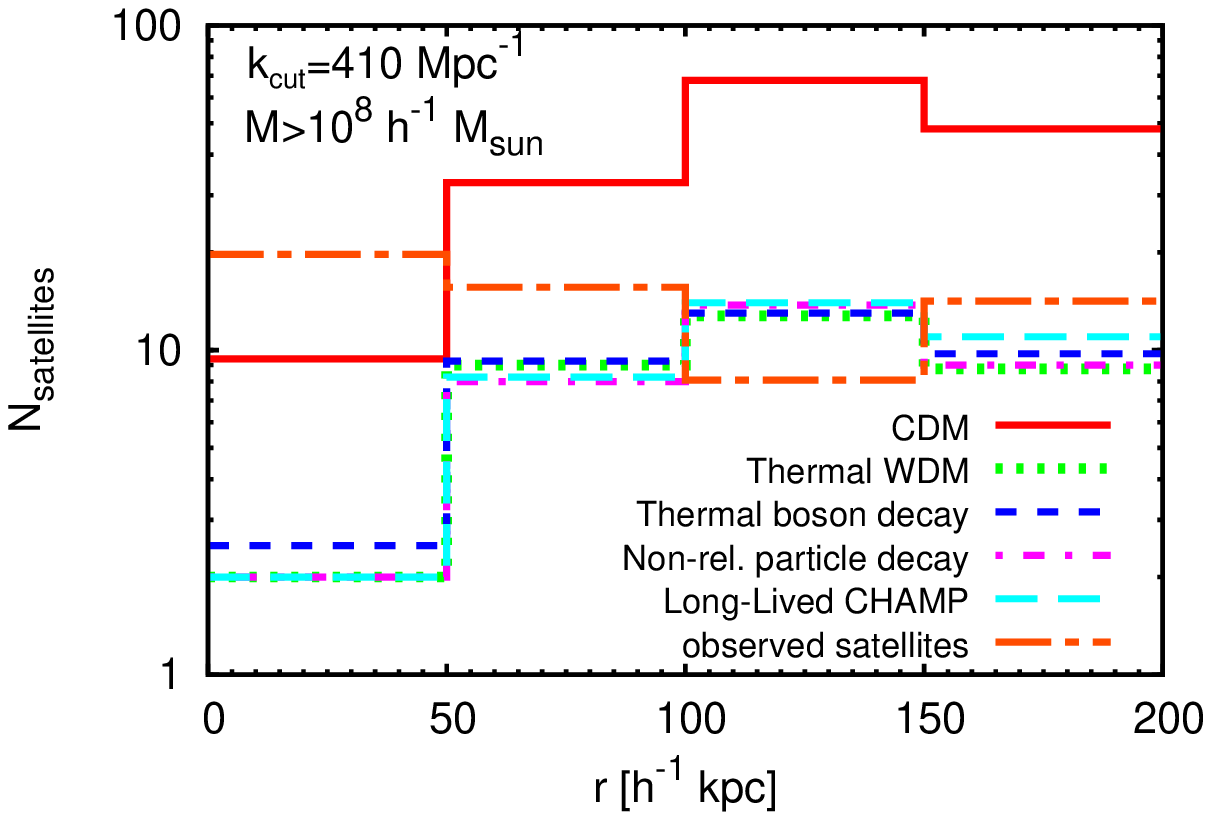}
 \end{center}
 \end{minipage}
 \begin{minipage}{.49\linewidth}
 \begin{center}
 \includegraphics[width=\linewidth]{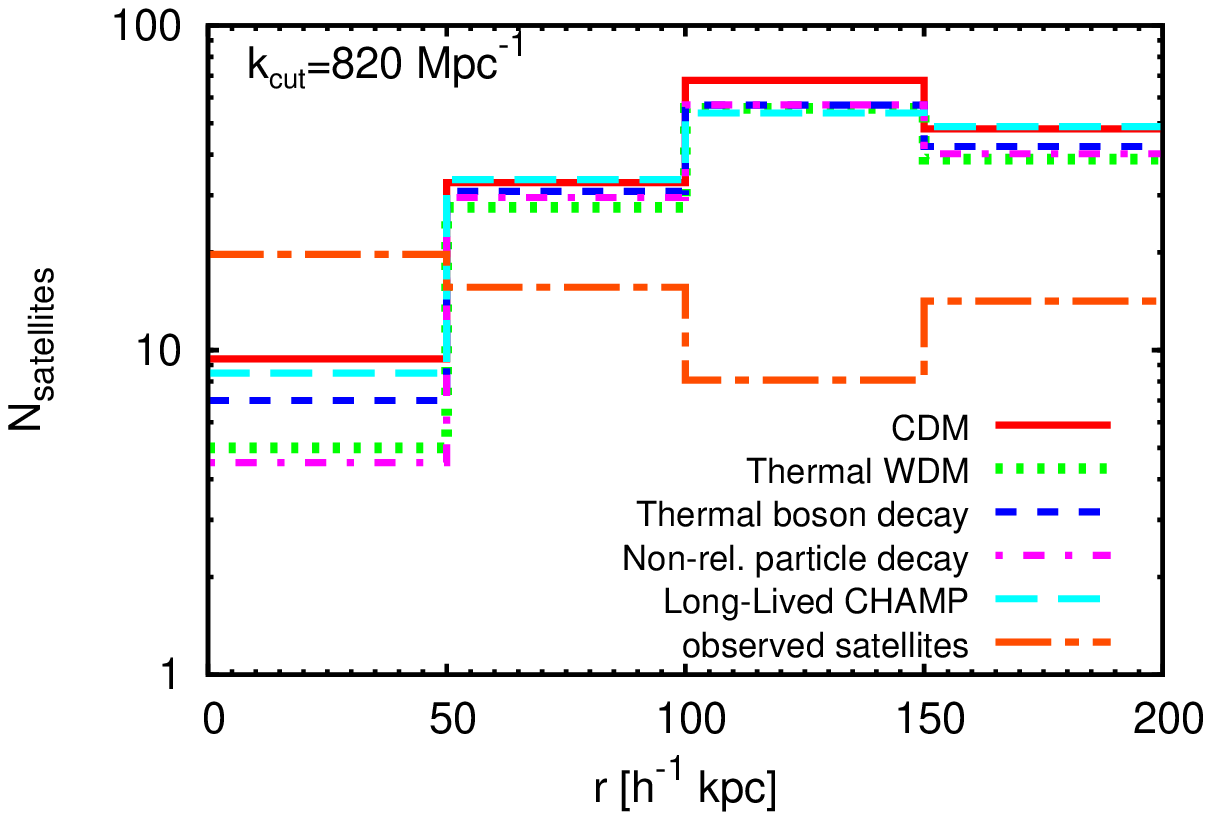}
 \end{center}
 \end{minipage}
 \caption{\sl \small
 Comparison of the averaged radial distributions in the CDM model, in the three WDM models and in the Long-Lived CHAMP model for $k_{\rm cut}\simeq410\,{\rm Mpc}^{-1}$ (left panel) and for $k_{\rm cut} \simeq 820\,{\rm Mpc}^{-1}$ (right panel).
 For $k_{\rm cut}\simeq410\,{\rm Mpc}^{-1}$ (left panel), we have discarded small subhalos ($M_{\rm sub} < M_{\rm c} \simeq 10^8\,h^{-1}\,M_{\rm sun}$) as in Fig.\,\ref{fig:twdm8radialdistirbution}.
}
\label{fig:compareradaildistribution}
\end{figure}

Let us now examine closely the similarity of the three WDM models 
and the Long-Lived CHAMP model when the characteristic cut-off scale $k_{\rm cut}$ is kept the same.
We compare the averaged radial distributions of the subhalos in these models 
for $k_{\rm cut}\simeq410\,{\rm Mpc}^{-1}$ (again after discarding small subhalos)
and for $k_{\rm cut}\simeq810\,{\rm Mpc}^{-1}$ in Fig.\,\ref{fig:compareradaildistribution}.
From Fig.\,\ref{fig:massfunction}, Fig.\,\ref{fig:nomsubcum} and Fig.\,\ref{fig:compareradaildistribution}, 
we conclude that the similar cut-off scale in the linear matter power spectra yields also
similar halo and subhalo abundances and radial distributions. 

\section{Summary}
\label{sec:summary}
In this paper, we study the formation of non-linear objects in three WDM models and 
in a Long-Lived CHAMP model.
We calculate the time evolution of the matter density fluctuations in the linear 
evolution regime by suitably modifying 
the public software {\tt CAMB}.
By using the obtained linear matter power spectra as initial conditions, we also 
perform large cosmological $N$-body simulations.
The results are summarized as follows.

First, the comoving Jeans scale at the matter-radiation equality characterizes 
the linear matter power spectra of WDM models well.
In the three WDM models motivated by particle physics, WDM particles are produced 
in different ways, but the linear matter power spectra with the 
same Jeans scale are very similar except for some difference at the damping tail at large $k$. 
We also consider a Long-Lived CHAMP model which has been suggested to yield a
cut-off of the matter power spectrum through the ``acoustic damping''.
The cut-off scale of the matter power spectrum in the Long-Lived CHAMP model is determined 
by the comoving horizon scale when CHAMP decays.
We empirically find the correspondence of the cut-off scales $k_{\rm cut}$ (see Eq.\,(\ref{eq:kcut})) between the three WDM models and the Long-Lived CHAMP model.

By performing large cosmological $N$-body simulations, we compare the abundances of nonlinear 
halos and subhalos and the radial distributions of the subhalos in Milky Way-size halos.
The three WDM models and the Long-Lived CHAMP model produce very similar halo and subhalo mass 
functions and radial distributions if $k_{\rm cut}$ is kept the same.
Therefore, we conclude that $k_{\rm cut}$ determines the clustering property of WDM 
and Long-Lived CHAMP well in both linear and non-linear growth of the matter density.

One might naively guess that our simulation results for small non-linear objects 
may be compromised by numerical effects.
However, our conclusions are drawn after discarding small objects that are likely
numerical artifacts.
We also compare the subhalo radial distributions in Milky Way-size halos with that of 
the observed satellites.
We find that the WDM models and the Long-Lived CHAMP model are broadly consistent with the observation
when they have $k_{\rm cut} \sim 50-800\,{\rm Mpc}^{-1}$. 
This cut-off scale corresponds to $m_{3/2} \sim 2-16\,{\rm keV}$ for the thermally-produced 
gravitino WDM and $\tau_{\rm Ch} \sim 0.01-2.5\,{\rm yr}$ for the Long-Lived CHAMP.
Because there is significant variation of the subhalo abundance among 
host halos with different masses, as reported by \citet{2009ApJ...696.2115I}, 
it would be important to use a large sample of halos in order to address
the validity of the models in a statistically complete manner.

Our results have a further implication for particles physics. 
We clarified how to put constraints on a few parameters of particles physics 
models which provide a WDM candidate. 
By calculating and comparing two quantities, the relic density of dark matter and the comoving Jeans 
scale at the matter-radiation equality,
one can apply the reported constraints in a specific particle physics model 
(e.g. the mass of thermally-produced gravitino WDM or sterile neutrino WDM produced 
through the \citet{1994PhRvL..72...17D} mechanism) to virtually any model parameters of interest.

Finally, we note that the so-called astrophysical feedback processes 
are also thought to affect the abundance of luminous satellite galaxies\,\citep{2012arXiv1209.5394B}.
Unfortunately, whether or not and how 
the baryonic processes changes our simple understanding based on 
the cut-off scale $k_{\rm cut} \sim 50-800\,{\rm Mpc}^{-1}$ 
is unclear due to the complexity of the baryonic processes.
Regarding ``missing satellite problem", both WDM and Long-Lived CHAMP models and the 
baryon feedback may explain the small number of the observed luminous satellites.
The degeneracy can be resolved, in principle, by the direct probes of non-luminous small objects, 
by, for instance, future submillilensing surveys\,\citep{2006PhLB..643..141H}.

\section{Acknowledgement}
This work is supported in part by JSPS Research Fellowship for Young Scientists (A.K.), by Grant-in-Aid
for the Ministry of Education, Culture, Sports, Science and Technology, Government of Japan Nos. 21111006,
22244030, 23540327 (K.K.), 23740195 (T.T.) and by World Premier International Research Center Initiative, MEXT, Japan.
Numerical computation in this work was carried out in part at the Yukawa Institute Computer Facility.

\clearpage









\end{document}